\documentclass[a4paper,11pt]{article}
\pdfoutput=1 

\usepackage{jinstpub} 

\usepackage{lineno}

\newlength{\figheight}
\title{The Silicon Vertex Detector of the Belle~II Experiment}

\author[a,1]{G.~Dujany\note{Speaker}}
\author[b]{K.~Adamczyk}
\author[c]{L.~Aggarwal}
\author[d]{H.~Aihara}
\author[e]{T.~Aziz}
\author[b]{S.~Bacher}
\author[f]{S.~Bahinipati}
\author[g, h]{G.~Batignani}
\author[a]{J.~Baudot}
\author[i]{P.~K.~Behera}
\author[g, h]{S.~Bettarini}
\author[j]{T.~Bilka}
\author[b]{A.~Bozek}
\author[k]{F.~Buchsteiner}
\author[g, h]{G.~Casarosa}
\author[g, h]{L.~Corona}
\author[l]{T.~Czank}
\author[m]{S.~B.~Das}
\author[a]{C.~Finck}
\author[g, h]{F.~Forti}
\author[k]{M.~Friedl}
\author[n,o]{A.~Gabrielli}
\author[n,o]{E.~Ganiev}
\author[o]{B.~Gobbo}
\author[e]{S.~Halder}
\author[p,q]{K.~Hara}
\author[e]{S.~Hazra}
\author[l]{T.~Higuchi}
\author[k]{C.~Irmler}
\author[p,q]{A.~Ishikawa}
\author[r]{H.~B.~Jeon}
\author[n,o]{Y.~Jin}
\author[l]{C.~Joo}
\author[b]{M.~Kaleta}
\author[e]{A.~B.~Kaliyar}
\author[j]{J.~Kandra}
\author[l]{K.~H.~Kang}
\author[b]{P.~Kapusta}
\author[j]{P.~Kody\v{s}}
\author[p]{T.~Kohriki}
\author[m]{M.~Kumar}
\author[c]{R.~Kumar}
\author[l]{C.~La~Licata}
\author[m]{K.~Lalwani}
\author[s]{R.~Leboucher}
\author[r]{S.~C.~Lee}
\author[i]{J.~Libby}
\author[a]{L.~Martel}
\author[g, h]{L.~Massaccesi}
\author[e]{S.~N.~Mayekar}
\author[e]{G.~B.~Mohanty}
\author[l]{T.~Morii}
\author[p,q]{K.~R.~Nakamura}
\author[b]{Z.~Natkaniec}
\author[d]{Y.~Onuki}
\author[b]{W.~Ostrowicz}
\author[g, h]{A.~Paladino}
\author[g, h]{E.~Paoloni}
\author[r]{H.~Park}
\author[s]{L.~Polat}
\author[e]{K.~K.~Rao}
\author[a]{I.~Ripp-Baudot}
\author[g, h]{G.~Rizzo}
\author[e]{D.~Sahoo}
\author[k]{C.~Schwanda}
\author[s]{J.~Serrano}
\author[p]{J.~Suzuki}
\author[p,q]{S.~Tanaka}
\author[d]{H.~Tanigawa}
\author[k]{R.~Thalmeier}
\author[e]{R.~Tiwari}
\author[p,q]{T.~Tsuboyama}
\author[d]{Y.~Uematsu}
\author[b]{O.~Verbycka}
\author[n,o]{L.~Vitale}
\author[d]{K.~Wan}
\author[d]{Z.~Wang}
\author[t]{J.~Webb}
\author[a]{J.~Wiechczynski}
\author[k]{H.~Yin}
\author[s]{L.~Zani}
\affiliation[a]{IPHC, UMR 7178, Universit$\acute{e}$ de Strasbourg, CNRS, 67037
Strasbourg, France}
\affiliation[b]{H. Niewodniczanski Institute of Nuclear Physics, Krakow 31-342, Poland}
\affiliation[c]{Panjab University, Chandigarh 160014, India}
\affiliation[d]{Department of Physics, University of Tokyo, Tokyo 113-0033, Japan}
\affiliation[e]{Tata Institute of Fundamental Research, Mumbai 400005, India}
\affiliation[f]{Indian Institute of Technology Bhubaneswar, Satya Nagar, India}
\affiliation[g]{Dipartimento di Fisica, Universit\`{a} di Pisa, I-56127 Pisa, Italy}
\affiliation[h]{INFN Sezione di Pisa, I-56127 Pisa, Italy}
\affiliation[i]{Indian Institute of Technology Madras, Chennai 600036, India}
\affiliation[j]{Faculty of Mathematics and Physics, Charles University, 121 16 Prague, Czech Republic}
\affiliation[k]{Institute of High Energy Physics, Austrian Academy of Sciences, 1050 Vienna, Austria}
\affiliation[l]{Kavli Institute for the Physics and Mathematics of the Universe (WPI), University of Tokyo, Kashiwa 277-8583, Japan}
\affiliation[m]{Malaviya National Institute of Technology Jaipur, Jaipur 302017, India}
\affiliation[n]{Dipartimento di Fisica, Universit\`{a} di Trieste, I-34127 Trieste, Italy}
\affiliation[o]{INFN Sezione di Trieste, I-34127 Trieste, Italy}
\affiliation[p]{High Energy Accelerator Research Organization (KEK), Tsukuba 305-0801, Japan}
\affiliation[q]{The Graduate University for Advanced Studies (SOKENDAI), Hayama 240-0193, Japan}
\affiliation[r]{Department of Physics, Kyungpook National University, Daegu 41566, Korea}
\affiliation[s]{Aix Marseille Universit$\acute{e}$ , CNRS/IN2P3, CPPM, 13288 Marseille, France}
\affiliation[t]{School of Physics, University of Melbourne, Melbourne, Victoria 3010, Australia}

\emailAdd{giulio.dujany@iphc.cnrs.fr}

\abstract{In 2019 the Belle~II experiment started data taking at the asymmetric
  SuperKEKB collider (KEK, Japan) operating at the Y(4S) resonance. Belle~II
  will search for new physics beyond the Standard Model by collecting an
  integrated luminosity of 50~ab$^{-1}$. The silicon vertex detector (SVD),
  consisting of four layers of double-sided silicon strip sensors, is one of the
  two vertex sub-detectors. The SVD extrapolates the tracks to the inner pixel
  detector (PXD) with enough precision to correctly identify hits in the PXD
  belonging to the track. In addition the SVD has standalone tracking capability
  and utilizes ionization to enhance particle identification in the low
  momentum region. The SVD is operating reliably and with high efficiency,
  despite exposure to the harsh beam background of the highest peak-luminosity
  collider ever built. High signal-to-noise ratio and hit efficiency have been
  measured, as well as the spatial resolution; all these quantities show
  excellent stability over time. Data-simulation agreement on cluster properties
  has recently been improved through a careful tuning of the simulation. The
  precise hit-time resolution can be exploited to reject out-of-time hits
  induced by beam background, which will make the SVD more robust against higher
  levels of background. During the first three years of running, radiation
  damage effects on strip noise, sensor currents and depletion voltage have been
  observed, as well as some coupling capacitor failure due to intense radiation
  bursts. None of these effects cause significant degradation in the detector
  performance.}

\collaboration[c]{(Belle~II SVD collaboration)}

\proceeding{12$^{\text{th}}$ International Conference on Position Sensitive Detectors\\
12-17 September, 2021\\
  Birmingham}

\begin{document}
\toccontinuoustrue
\maketitle
\flushbottom
\thispagestyle{empty}
\newpage
\pagestyle{myplain}\pagenumbering{arabic}

\section{Introduction}
\label{sec:intro}
Belle~II~\cite{BelleII} is a high energy physics experiment at the intensity
frontier searching for physics beyond the Standard Model in rare $B$ meson,
charm and tau decays. Since 2019, it has been operating at the $e^+e^-$
asymmetric collider SuperKEKB~\cite{SuperKEKB} at Tsukuba in Japan. This
collider operates primarily at the centre-of-mass energy corresponding to the
$\Upsilon(4S)$ mass (10.58~GeV) and it is designed to reach an instantaneous
luminosity of $6 \times 10^{35}$~cm$^{-2}$~s$^{-1}$ and to deliver a
$50$~ab$^{-1}$ dataset. It already holds the world luminosity record of $3.1
  \times 10^{34}$~cm$^{-2}$~s$^{-1}$. Essential features required to reach
Belle~II physics goals are the precise determination of the decay vertices,
accurate and efficient tracking capabilities, including low-momentum particles, and
the capability to distinguish between the different kinds of charged~particles.

\begin{figure}[htb]
  \centering
  \includegraphics[width=0.7\columnwidth]{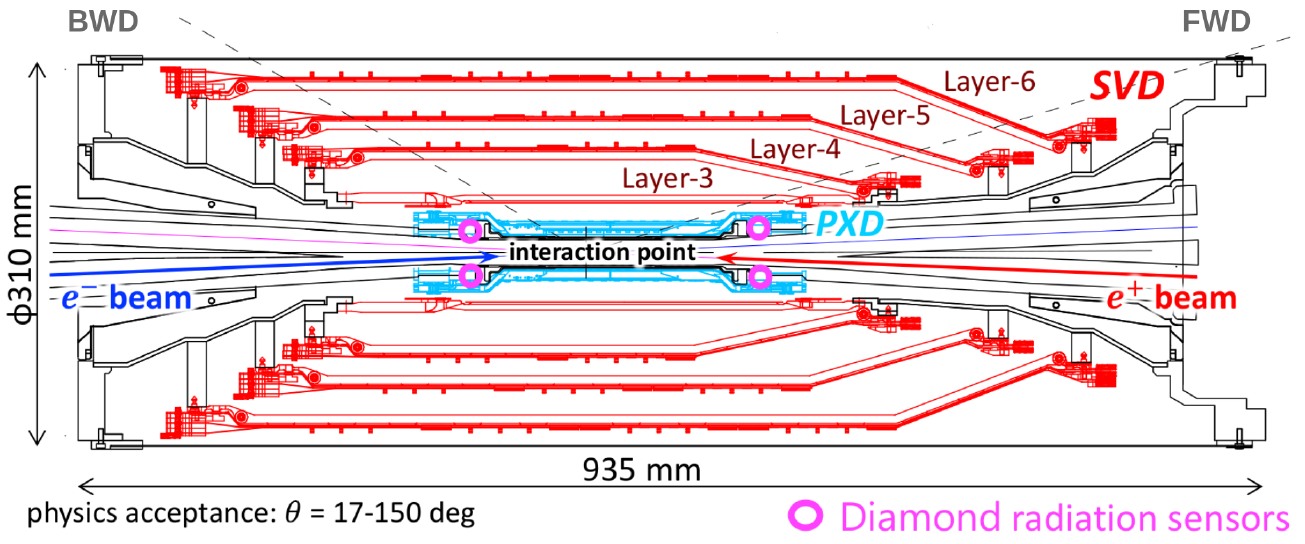}
  \caption{\label{fig:layout}  The cross-section view of the Belle~II vertex detector: SVD and PXD are represented with red and light blue colour,
    respectively. The pink circles indicate the positions of the diamond sensors installed on the beam pipe.
  }
\end{figure}

The Belle~II experiment is an upgrade of the Belle the detector. Its design has
been conceived to allow similar or better performance than Belle in an harsher
environment characterised by an higher beam background and Lorentz
boost reduced from $\beta\gamma = 0.4$ to $\beta\gamma = 0.3$. The vertex detector of
Belle~II is designed to provide a better resolution than the Belle silicon
vertex detector~\cite{Belle} in order to compensate for the reduced Lorentz
boost. This improvement is achieved with more precise point resolution, a
reduced inner radius and a low material budget. Moreover it must operate in a
high background environment with a hit rate of 20~(3)~MHz/cm$^{2}$ and an
integrated yearly dose of 2~(0.2)~Mrad at a radius of 14~(40)~mm.
Figure~\ref{fig:layout} shows the cross-section view of the Belle~II
vertex detector. It is composed of two layers of DEPFET pixel sensors (PXD), with
the innermost layer at 1.4~cm from the interaction point, and four layers of
double-sided silicon strip sensors (SVD).

\section{The Belle II silicon strip vertex detector}
\label{sec:svd}
The SVD detector~\cite{SVD} consists of four layers numbered 3, 4, 5 and 6 from lower to
higher radius, which are composed of seven, ten, twelve and sixteen
ladders with two, three, four and five sensors, respectively. Its material budget is
0.7\% of the radiation length $X_0$ per layer on average. Diamond sensors are
installed on the beam pipe for radiation monitoring and to trigger fast beam
aborts~\cite{diamonds}. The SVD detector has three specific roles: firstly it
allows to extrapolate the tracks toward the PXD; this is essential for the
reconstruction of decay vertices and also to define regions of interest to
reduce the PXD data transfer rate. Secondly it provides stand-alone tracking for low
momentum tracks and finally it contributes to the particle identification by
providing the information about the ionization energy loss.

The SVD detector is composed of 172 double-sided silicon strip sensors covering
an area of 1.2~m$^2$ for a total of 224 thousand strips. Each sensor is based on
an N-type bulk about 320~$\mu$m thick. The two opposite sides are implanted with
P- and N-doped sensitive strips. The metal readout strips are AC coupled and are
placed on top of the implanted strips, separated by a dielectric SiO$_2$. Each
readout channel is connected to a single strip. Between every two readout strips
there is a floating strip that is not read out. The two sensor sides are
referred to as the u/P and the v/N sides and have strips parallel and orthogonal
to the beam direction, respectively. Layer 3 is equipped with $40$~mm $\times
125$~mm sensors with a pitch of 50~(160)~$\mu$m for the u/P~(v/N) side. Layers
4, 5 and 6 are equipped with $60$~mm $\times 125$~mm rectangular sensors with a
pitch of 75~(240)~$\mu$m in u/P (v/N) and with trapezoidal slanted sensors
$126$~mm long, with a width varying from 41 to 61~mm and a pitch varying from 50
to 75~$\mu$m in u/P and a 240~$\mu$m pitch in v/N. The sensor full-depletion
voltage is between 20 and 60~V and the operation voltage is 100~V. The
rectangular sensors have been produced by Hamamatsu Photonics while Micron
Semiconductor produced the trapezoidal slanted ones.

The SVD front-end readout ASIC is the APV25 chip~\cite{APV25}. It was originally
developed for the CMS silicon tracker. It has 128 channel inputs per chip, a
short shaping time of 50~ns, its power consumption is 0.4~W per chip and it can
tolerate more than 100~Mrad of radiation dose. It operates in a multi-peak mode
at 32~MHz, and not at the bunch crossing frequency, which is about eight
times larger. To reconstruct the output waveform of each channel, six subsequent
analog samples are recorded. A mixed three/six acquisition mode has also been
prepared to reduce dead time, data size and occupancy at higher luminosity. The
readout chips of the middle sensors are directly installed on one side of the
sensor to minimise the signal propagation length and reduce the capacitance and
the noise. These chips are thinned to 100~$\mu$m to reduce the material budget
and are cooled down using bi-phase CO$_2$ at -20$^\circ$~C. A wrapped flex also
allows the read out of the side of the sensor opposite to the chip position.

\section{Operation and performance}
\label{sec:performance}
\begin{figure}[htb]
  \centering
  \includegraphics[width=0.48\columnwidth]{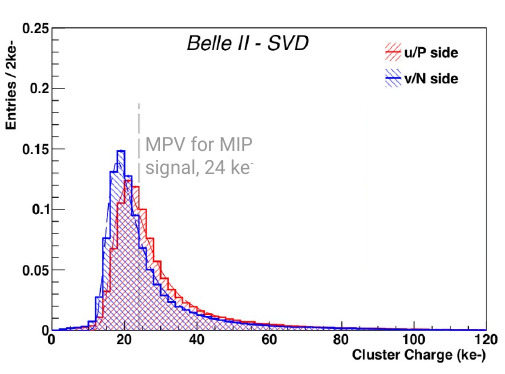}
  \includegraphics[width=0.48\columnwidth]{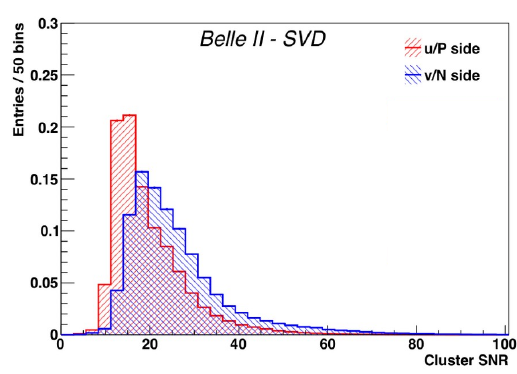}
  \caption{\label{fig:SNR} Distributions of (\textit{left}) the cluster charge
    normalised with the track path length and scaled to the sensor thickness and
    (\textit{right}) signal-to-noise ratio (SNR) for the more backward sensors
    of layer 3, summed over all $\phi$ ladders. The red and the blue
    distributions correspond to the u/P and v/N side respectively. The dashed
    grey line in correspondence to 24~k$e^-$ in the left plot represents the
    expected most probable value (MPV) for a minimal ionising particle~(MIP).
    Data for a typical run with colliding beams in 2019 corresponding
    to 14~pb$^{-1}$ are shown.}
\end{figure}

The SVD detector has been installed in 2018 and has been operated since 2019.
The operation has been reliable and smooth. The total fraction of masked strips
is about $1\%$, moreover the average sensor hit efficiency is greater than
$99\%$ and stable with time. The collected signal charge, normalised for the
length of traversed sensor thickness, is similar in all sensors and matches the
expectations. In the u/P side the charge is in agreement with the expectations
from a minimal ionising particle taking into account the uncertainty in the
APV25 gain calibration ($\sim 15\%$). In the v/N side there is a $10\%-30\%$
signal loss due to the large pitch and the presence of the floating strip. A
very good signal-to-noise ratio is observed in all 172 sensors with a most
probable value between 13~to~30. The u/P side presents a larger noise due to the
longer strip length and the larger inter-strip capacitance. The signal instead
depends strongly by the sensor position due to the track incidence angle.
Figure~\ref{fig:SNR} shows the collected signal charge, normalised for the
length of traversed sensor thickness and the signal-to-noise ratio for the more
backward sensors of layer 3, summed over all $\phi$ ladders.

\begin{figure}[htb]
  \centering
  \includegraphics[height=\figheight]{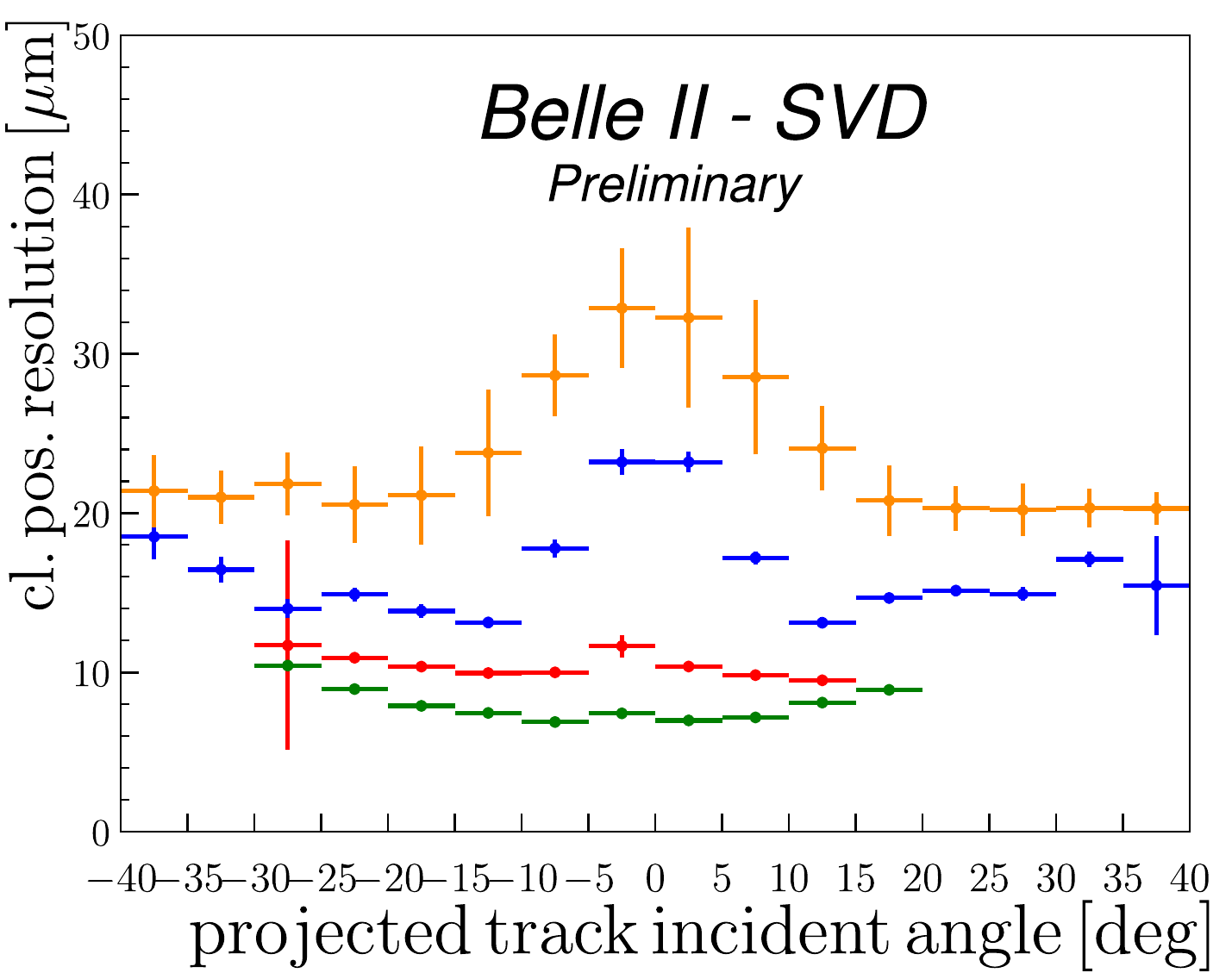}
  \includegraphics[height=\figheight]{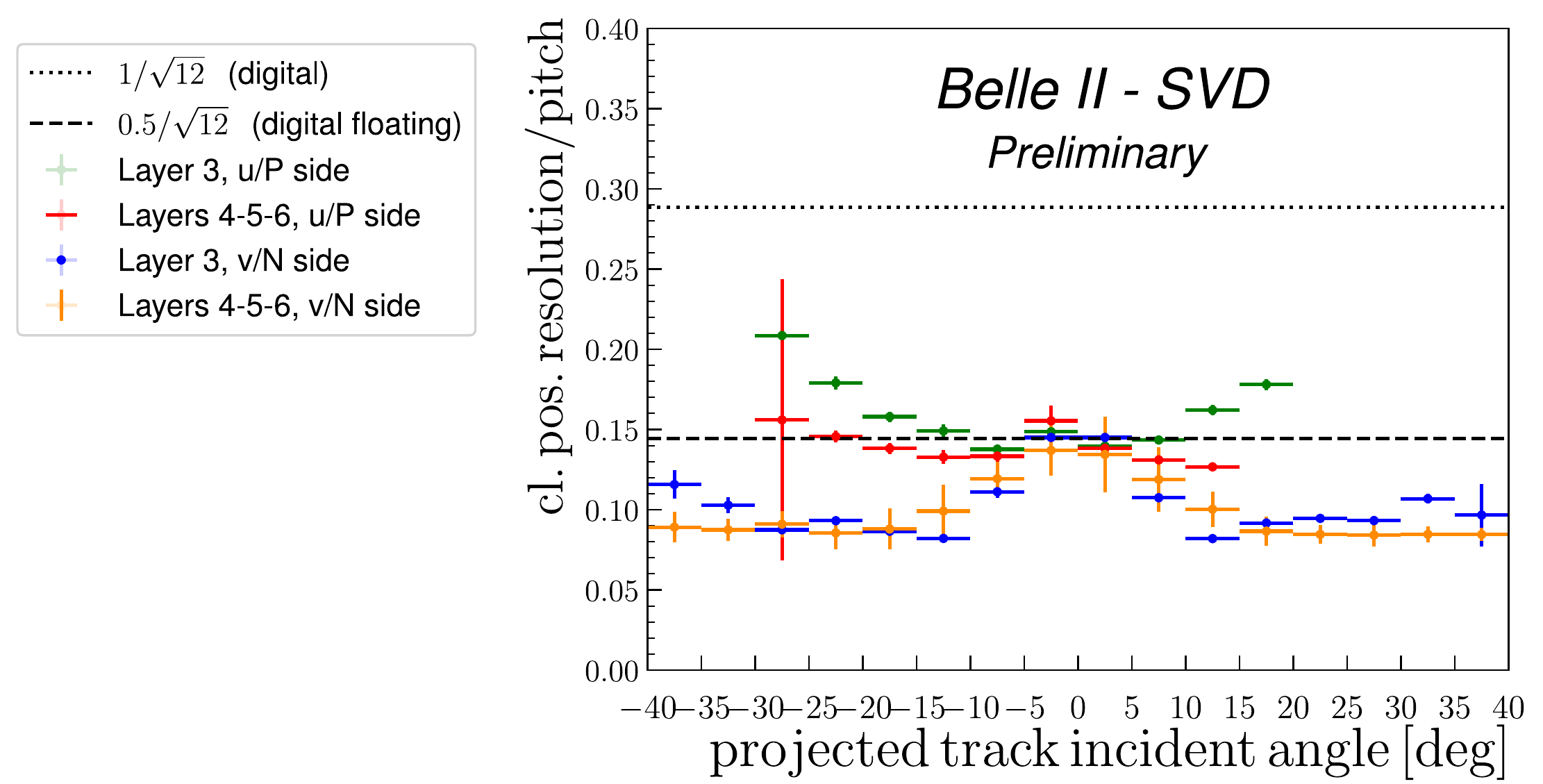}
  \caption{\label{fig:reso} (\textit{left}) Cluster position resolution and
    (\textit{right}) cluster position resolution divided by the pitch as a
    function of the angle between the direction orthogonal to the sensor plane
    and the projection of the incident track into the plane orthogonal to the
    strip. These results have been obtained using $e^+e^- \to \mu^+ \mu^-$
    events collected in 2020 corresponding to 98~pb$^{-1}$.}
\end{figure}

The cluster position resolution has been measured using
$e^+e^- \to \mu^+ \mu^-$ events. It is estimated from the residuals of the
cluster position with respect to the unbiased track intercept after the effect of
the track extrapolation error has been subtracted. The measured resolutions are in
agreement with the expectations from the pitch and are
approximately 9~$\mu$m for layer 3 u/P side, 11~$\mu$m for layer 4, 5 and 6 u/P
side, 20~$\mu$m for layer 3 v/N side and 25~$\mu$m for layer 4, 5 and 6 v/N
side. Figure~\ref{fig:reso} shows the distributions, with
respect to the projected track incident angle, of the resolution and of the
resolution normalised to the pitch.
The hit time resolution with respect to the event time provided by the Belle~II
central drift chamber is approximately 2.9~ns for the u/P side and 2.4~ns for
the v/N side. This excellent hit time resolution can be exploited to efficiently
reject off-time hits coming from the beam background. Preliminary studies show
that it is possible to reject about half of the background hits keeping
more than $99\%$ of the signal ones. This information, not yet used, will be
exploited at higher luminosities and background levels.

\section{Beam background and radiation effects}
\label{sec:background}
Beam background produces radiation damage in the sensors and increases the SVD
hit occupancy. A higher hit occupancy degrades the tracking performance by
increasing the number of clusters that can be erroneously associated to tracks
or even create fake tracks in the reconstruction. The present occupancy limit,
set by performance studies, is about $3\%$ in layer~3. This limit could be
approximately doubled by exploiting the hit time information to reject off-time
background hits. With the current luminosity the average hit occupancy in layer
3 is less than $0.5\%$. It however reaches $3\%$ at the design luminosity of $8
  \times 10^{35}$~cm$^{-2}$~s$^{-1}$ according to beam background projections based on
scaling the simulation with a data-simulation ratio. This corresponds to a dose
of 0.2~Mrad per year and a 1~MeV neutron equivalent fluence of $5\times
  10^{11}$~n$_{eq}$/cm$^{2}$ per year. These long term beam background
extrapolations are affected by large uncertainties related to the collimation
and the injection background that is not yet included. The integrated dose on
the SVD sensors is evaluated exploiting the correlation between the measured
occupancy and the dose rate measured in the diamond sensors~\cite{diamonds}. This estimate is
based on several assumptions and has an uncertainty of about $50\%$.
The mid plane of layer
3 is the detector region the most exposed to radiation. The
estimated dose collected by this region over the past two and a half years is 70~krad.
This corresponds to a fluence of
$1.6 \times 10^{11}$~n$_{eq}$/cm$^2$, where the ratio of $2.3\times
  10^9$~n$_{eq}$/cm$^2$/krad between dose and n$_{eq}$ fluence  was estimated from
simulation.

The first effects of radiation damage are already visible. The increase of
leakage currents are proportional to the sensor bulk damage due to non-ionising
energy loss. This damage, according to the NIEL hypothesis~\cite{niel}, is
proportional to the 1~MeV neutron equivalent fluence, which in turn is
proportional to the integrated dose. This linear correlation between leakage
current and integrated dose is observed for all sensors with slopes in the range
2 to 5~$\mu$A/cm$^{2}$/Mrad. These values are of the same order of magnitude as
those observed by the BaBar experiment: 1~$\mu$A/cm$^{2}$/Mrad at
20$^\circ$C~\cite{babar}. In the SVD the leakage current contribution to the
noise is highly suppressed by the short shaping time of the APV25
chips. Therefore, it will not affect significantly the total strip noise even
after ten years at design luminosity, corresponding to about 2~Mrad of dose with
present background extrapolation. It however will  become comparable with present
noise after about ten Mrad. Sensor bulk damage due to non-ionising energy loss
can also change the depletion voltage. The depletion voltage can be monitored by
measuring the v/N side noise as a function of the operation voltage: since the
substrate of the sensor is N-type, the v/N-side strips are completely isolated,
and hence their noise drops, at full depletion. No change in the full depletion
voltage has yet been observed, in agreement with the low integrated 1~MeV
neutron equivalent fluence accumulated. Another radiation effect is the increase
of strip noise due to the ionising energy loss that causes fixed oxide charges
in the SiO$_2$ layer increasing the inter-strip capacitance. The amount of fixed
oxide charges and hence the noise increase is expected to saturate. A noise
increase of 20-25\% has been observed in layer 3. The saturation has been
reached in the v/N side and is being reached in the u/P side after 70~krad. No
impact on performances is observed due to this increase of the strip noise.

\section{Conclusion}
\label{sec:conclusion}

The Belle~II SVD detector has been taking data since March 2019 smoothly and
reliably, showing excellent performance in agreement with the expectations. The
first effects of radiation damage have been observed at the expected level and
they are not affecting the performance. The detector is ready to cope with the
increased beam background expected during future running.

\acknowledgments

This project has received funding from the European Union's Horizon 2020
research and innovation programme under the Marie Sklodowska-Curie grant
agreements No 644294 and 822070. This work is supported by MEXT, WPI, and JSPS
(Japan); ARC (Australia); BMBWF (Austria); MSMT (Czechia); L’Institut National
de Physique Nucléaire et de Physique des Particules (IN2P3) du CNRS (France);
AIDA-2020 (Germany); DAE and DST (India); INFN (Italy); NRF and RSRI (Korea);
and MNiSW (Poland).

\end{document}